% ...

\documentclass[a4paper,fleqn,usenatbib]{mnras}

\usepackage[dvips]{graphicx}
\usepackage{colortbl,color}
\usepackage{amssymb,txfonts}
\usepackage[T1]{fontenc}
\usepackage{ae,aecompl}
\def\ee{\end{equation}}
\def\be{\begin{equation}}

\newcommand{\mbh}{M_{\rm BH}}
\newcommand{\msun}{{\rm M}_{\sun}}
\newcommand{\ledd}{L_{{\rm Edd}}}
\newcommand{\mdot}{\dot M}
\newcommand{\medd}{{\dot M_{\rm Edd}}}
\newcommand{\ergs}{{\rm erg\,s^{-1}}}

\newcommand{\kev}{{\rm keV}}

\title[Radio/X-ray correlation of BHBs]{Interpreting the  radio/X-ray correlation of black hole X-ray binaries based on the accretion--jet model}

\author[F. G. Xie and F. Yuan]{Fu-Guo Xie\thanks{E-mail: fgxie@shao.ac.cn} and
Feng Yuan\\
Key Laboratory for Research in Galaxies and Cosmology, Shanghai Astronomical Observatory, \\
Chinese Academy of Sciences, 80 Nandan Road, Shanghai 200030, China\\
}

\begin{document}
\label{firstpage}
\pagerange{\pageref{firstpage}--\pageref{lastpage}}
\maketitle

\begin{abstract}
Two types of correlations between the radio and X-ray luminosities ($L_{\rm R}$ and $L_{\rm X}$) have been found  in black hole X-ray binaries. For some sources, they follow the `original' type of correlation which is described by a single power-law. Later it was found that some other sources follow a different correlation consisting of three power-law branches, with each branch having different power-law indexes. In this work, we explain these two types of correlation under the coupled accretion--jet model. We attribute the difference between these two types of sources to the different value of viscosity parameter $\alpha$. One possible reason for different $\alpha$ is the different configuration of magnetic field in the accretion material coming from the companion stars. For the `single power-law' sources, their $\alpha$ is high; so their accretion is always in the mode of advection-dominated accretion flow (ADAF) for the whole range of X-ray luminosity. For those `hybrid power-law' sources, the value of  $\alpha$ is small so their accretion mode changes from an ADAF to a luminous hot accretion flow, and eventually to two-phase  accretion as the accretion rate increases. Because the dependence of radiative efficiency on the mass accretion rate is different for these three accretion modes, different power-law indexes in the $L_{\rm R} - L_{\rm X}$ correlation are expected. Constraints on the ratio of the mass loss rate into the jet and the mass accretion rate in the accretion flow are obtained, which can be tested in future by radiative magnetohydrodynamic numerical simulations of jet formation.

\end{abstract}


\section{Introduction}
A strong correlation between the radio luminosity (e.g., at 8.6 GHz; hereafter $L_{\rm R}$) and X-ray luminosity (e.g., at 3-9 keV; hereafter $L_{\rm X}$) has been found in black hole (BH) X-ray binaries (BHBs) in their hard state (e.g., Corbel et al. 2000, 2003, 2013; Gallo et al. 2003; but see Xue \& Cui 2007). This correlation can be well described by a single power-law form, $L_R\propto L_X^p$, with the index $p\approx 0.6$ (Corbel et al. 2013). We call this the `original' radio/X-ray correlation. The correlation was  extended to including low-luminosity active galactic nuclei (AGNs) by considering the effect of the BH mass (e.g., Merloni et al. 2003; Falcke et al. 2004; K\"ording et al. 2006; Wang et al. 2006; Li et al. 2008; G\"ultekin et al. 2009; Younes et al. 2012).

However, it was soon found that some sources do not follow this single power-law correlation.
Rather, they follow a `hybrid' correlation, i.e. the correlation index varies in different regimes of luminosity. Among them H1743-322 is a prototype, thanks to the frequent outbursts it undergoes and the sufficient large coverage in its $L_{\rm X}$ during the outbursts. The radio/X-ray correlation index of this source shows a transition from $p\approx1.4$ at the bright $L_{\rm X}$ regime, through $p\sim 0.2$ (almost a flat correlation) at moderate $L_{\rm X}$ regime, eventually to the `original' $p\approx 0.6$ at the weak $L_{\rm X}$ regime (Coriat et al. 2011). Such kind of hybrid correlation was also discovered, but with fewer data points and/or narrower coverage in $L_{\rm X}$, in MAXI J1659-152 (Jonker et al. 2012) and XTE J1752-223 (Ratti et al. 2012; Brocksopp et al. 2013). Statistical analysis of 18 BHBs by Gallo et al. (2012) also shows that, in addition to the original $p\approx0.6$ correlation, another correlation with $p\approx 0.98$ exists at the bright $L_{\rm X}$ regime. All these observational results are summarized in Fig.\ \ref{fig:rx}.

Theoretically the `original' $p\approx 0.6$ correlation has been explained in the framework of the coupled accretion--jet model (Heinz \& Sunyaev 2003; Merloni et al. 2003; Heinz 2004; Yuan \& Cui 2005). In this model, the radio radiation comes from the synchrotron emission from the jet, while X-ray from the Comptonization emission from the hot accretion flow. In this case, the correlation index $p$ is mainly determined by three factors, namely the dependence of $L_{\rm R}$ on the mass-loss rate in the jet $\mdot_{\rm jet}$ (i.e., the radio radiative efficiency of jet), the dependence of $L_{\rm X}$ on the mass accretion rate $\mdot_{\rm in}$ (i.e., the X-ray radiative efficiency of the accretion flow), and finally the fraction of accreting material that enters into the jet $\eta_{\rm jet}$ (see Eq. 1 below for definition) as a function of accretion rate.

In addition to explaining the observed `original' correlation, two predictions are made by Yuan \& Cui (2005) based on the accretion--jet model by extrapolating $\eta_{\rm jet}$ to lower accretion rates. One is that when the $L_{\rm X}$ $L_{\rm X}$ is below a critical value, $L_{\rm X, crit}\sim 10^{-5}-10^{-6}~\ledd$ ($\ledd$ is the Eddington luminosity), the X-ray radiation from the jet will exceed that from the accretion flow. Physically, this is because the jet also emits X-ray radiation and this emission is less sensitive to the accretion rate compared with the radiation from the accretion flow; thus below a very low accretion rate, the radiation from the jet will catch up with that from the accretion flow and even become a dominant contributor. This explains why the X-ray emission of several very low luminosity sources such as M87 is dominated by the jet (Wilson \& Yang 2002; Yuan, Yu \& Ho 2009). The second prediction is that the radio/X-ray correlation should become steeper, i.e. $p\approx 1.23$ (Yuan \& Cui 2005, see also Heinz 2004). Both predictions have been confirmed by later works, both observational and theoretical (Pellegrini et al. 2007; Wu et al. 2007; Pszota et al. 2008; Wrobel et al. 2008; Yuan et al. 2009; de Gasperin et al. 2011; Younes et al. 2012). For example, it is found that the data from all available low-luminosity AGNs satisfying $L_{\rm X}\la L_{\rm X, crit}$ follows $L_{\rm R}\propto L_{\rm X}^{1.22}$, in perfect agreement with the Yuan \& CUi (2005) prediction. In the case of BH X-ray binaries, however, the answer is less clear due to two reasons. One is that the data points of BHBs in quiescent states are still very limited. Another is that the data quality of radio observations at such low X-ray luminosities is still very poor (Yuan \& Narayan 2014). For example, the radio detection of XTE J1118+480 at its quiescent state is marginal, only at 3$\sigma$ level (Gallo et al. 2014).

In this work, we focus on relatively luminous observations. The question we want to address is whether we can also understand the `hybrid' correlation based on the accretion--jet model. Before we begin our study, we would like to mention several models proposed recently. Meyer-Hofmeister \& Meyer (2014, see also Cao, Wu \& Dong 2014; Huang, Wu \& Wang 2014 and Qiao \& Liu 2015) propose that when accretion rate is high, a weak, cool disc will be formed in the innermost region of the hot accretion flow. This cool disc supplies additional seed photons for Compton scattering, which will significantly enhance the produced X-ray flux. They propose it can explain the $p\approx 1.4$ radio/X-ray correlation, i.e. the bright part of the `hybrid' correlation. However, they do not explain why there is also a flat $p\sim 0.2$ branch and why different sources follow different correlations. These questions will be addressed in this work.

This paper is organized as follows. In Section 2 we briefly review the accretion--jet model. We then calculate the radiation from both the hot accretion flow and jet to explain the correlations in Section 3. The last section (Section 4) devotes to a summary and discussion.

\section{The accretion--jet model}

There are three components in this model, i.e. an outer truncated Shakura--Sunyaev disc (Shakura \& Sunyaev 1973, SSD hereafter), an inner hot accretion flow, and a jet. The details of this model are described in Yuan, Cui \& Narayan (2005, hereafter YCN05).

\subsection{The jet model}
Recently many magnetohydrodynamical (MHD) numerical simulations have been performed to study the jet formation (see review in Yuan \& Narayan 2014). Most of these works focus on the `BZ-jet', which is formed by extracting the spin energy of the BH [e.g., Blandford \& Znajek 1977 (BZ); Tchekhovskoy et al. 2011]. It is a relativistic, Poynting flux-dominated jet. In addition to the BZ jet, numerical simulations have revealed another type of jet called `disc-jet', which is formed by extracting the rotation energy of the underlying disc. It is a quasi-relativistic, matter-dominated jet. The `disk-jet' can even exist around a non-rotating BH since it is powered by the rotation of the accretion flow (e.g., Lynden-Bell 2003; Kato et al. 2004; Hawley \& Krolik 2006; Ohsuga \& Mineshige 2011; Yuan et al. 2015).

The jet scenario adopted here seems to be more close to the `disk-jet' model, because the composition of the jet is assumed to be dominated by normal plasma, i.e. electrons and ions, coming from the underlying hot accretion flow. The existence of protons in the jet is supported by several observations of the AGN jets (see Sikora 2011 for a review): (1) the detection of circular polarization and Faraday rotation of the radio core. This is because the electron--positron plasma cannot generate any polarization (Park \& Blackman 2010); (2) the low-energy cutoff in the radio spectra (and also the electron energy distribution) of hotspots in radio-lobes, which is likely a consequence of dissipation of bulk kinetic energy in an electron--proton jet (Godfrey et al. 2009). We mainly follow Spada et al. (2001) to calculate the radiation from the jet\footnote{Recently Kumar \& Crumley (2015) calculate of the radiation from the `BZ-jet'.}. The half opening angle of the jet is assumed to be $\theta=0.1$, and the bulk Lorentz factor is fixed to $\Gamma_{\rm jet}=1.2$. Within the jet, internal shocks occur due to the collision of shells with different velocities. These shocks accelerate a fraction of the electrons ($\xi=0.01$) into a power-law energy distribution. Shock acceleration theory predicts that the power-law index $p_{\rm jet}$ of these non-thermal electrons to be $2<p_{\rm jet}<3$, and we set $p_{\rm jet}=2.14$ throughout this work. The steady-state energy distribution of the accelerated electrons is then self-consistently calculated, taking into account the effect of radiative cooling. The energy density of accelerated electrons and amplified magnetic field is determined by two parameters, $\epsilon_e$ and $\epsilon_B$, which describe the fraction of the shock energy that goes into electrons and magnetic fields, respectively. Theoretical study of relativistic collisionless
electron-proton plasma puts a constrain on $\epsilon_e$ and $\epsilon_B$, i.e. $\epsilon_e \sim \sqrt{\epsilon_B}$ (Medvedev 2006). We thus take the values of $\epsilon_e$ and $\epsilon_B$ to be 0.1 and 0.02, respectively. These values are also within the typical range obtained in the study of GRB afterglows (Medvedev 2006). We note that different values of $\epsilon_e$ and $\epsilon_B$ will only change the normalization, but not the slope, of the $\eta_{\rm jet}--\mdot_{\rm in}(5R_{\rm s})$ relationship (cf. equation \ref{fig:eta} below). We then calculate the synchrotron emission from these accelerated electrons.

Like any other jet models in literature, our jet model is phenomenological. Besides, we assume that all the jet parameters remain unchanged for different sources and accretion rates, which is certainly a strong assumption. This is partly because we still do not have good constraint on them. To conclude, the only free parameter in our jet model is its mass-loss rate $\mdot_{\rm jet}$. We note that the $L_{\rm R}$ is found to be a power-law function of $\mdot_{\rm jet}$ (see also Heinz \& Sunyaev 2003).

\subsection{Hot accretion flow model}

Depending on the accretion rate, two types of hot accretion flow exist. They are the advection-dominated accretion flow (ADAF; Narayan \& Yi 1994, 1995; Abramowicz et al. 1995) below certain critical accretion rate $\mdot_{\rm cr, ADAF}$ and the luminous hot accretion flow (LHAF; Yuan 2001) above it (see review by Yuan \& Narayan 2014). LHAF is thermally unstable (Yuan 2003). But if accretion rate is relatively low, $\mdot_{\rm in}\la \mdot_{\rm cr, LHAF}$, the growth time-scale of the thermal instability is larger  than the accretion time-scale, so the gas can remain hot throughout the flow. We call it type-I LHAF. Above $\mdot_{\rm cr, LHAF}$, the growth time-scale of the thermal instability is shorter than the accretion time-scale thus some cold dense clumps should be formed, embedded in the hot phase medium. We call it type-II LHAF or two-phase accretion flow.

We follow the standard approach to calculate the dynamical structure and the emitted spectrum of ADAFs and Type-I LHAF (e.g., YCN05). The $L_{\rm X}$ at $3-9~\kev$ band is then derived from the spectrum. For the two-phase accretion flow, optical/UV radiation from cold clumps will provide additional (likely dominant) seed photons for the inverse Compton scattering process, to generate the X-ray emission. The whole process is obviously complicated, depending on the detailed dynamics of the two-phases accretion flow such as the filling factor and temperature of clumps and so on. Following Yuan \& Zdziarski (2004), we simplify the problem by replacing the electron energy equation with the Compton $y$-parameter. With the assumption that the electron advection term is zero in the two-phase flow, we can calculate the radiative cooling rate. The bolometric luminosity can then be derived (Xie \& Yuan 2012). In order to derive $3-9~\kev$ $L_{\rm X}$ $L_{\rm X}$, we assume the spectrum to be a simple power-law with exponential cutoffs at both the high and low ends, i.e. $F_E\sim E^{1-\Gamma} \exp(-E_{\rm min}/E) \exp(-E/E_{\rm max})$, where the photon index $\Gamma$ is constrained by the Compton $y$-parameter. $E_{\rm min}$ is set arbitrarily to be $0.02~\kev$. $E_{\rm max}$ is numerically determined by the electron temperature, i.e. $E_{\rm max}=k T_{\rm e}$, at the location where most of the radiation comes out (Yuan \& Zdziarski 2004).

\begin{figure*}
\centerline{\includegraphics[width=15cm]{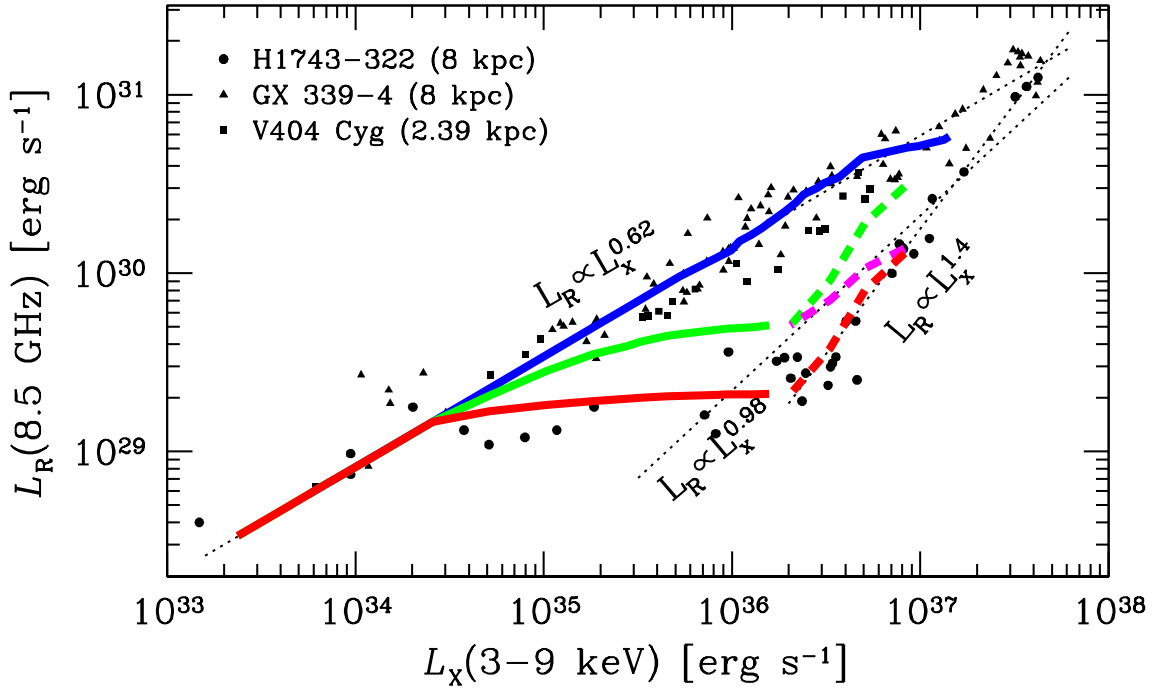}}
\caption{The radio/X-ray correlation of BH X-ray binaries. As labelled in the figure, the points are observational data from three representatives. The three dotted lines show the fitting results to the data. The lines with different colour show the results calculated from the accretion--jet model. The (blue, green, and red) solid lines represent the pure hot solutions (ADAF and type-I LHAF), while the (green, magenta, and red) dashed lines represent the result of two-phase accretion model. The blue line corresponds to $\alpha=0.6$, and others to $\alpha=0.06$. The corresponding $\eta_{\rm jet}--\mdot_{\rm in}$ relationship is shown in Fig.\ \ref{fig:eta}. Note that the curves are superimposed at $L_{\rm X}\la 2\times10^{34}\ergs$.}
\label{fig:rx}
\end{figure*}

\subsection{Basic parameters of the accretion--jet model}

In our calculations, the BH mass is fixed to $\mbh=10~\msun$. In addition to the accretion rate $\mdot_{\rm in}$ of the accretion flow, our accretion--jet model contains several basic parameters as described below.

Hot accretion flows are subject to strong wind (Yuan, Bu \& Wu 2012; Yuan et al. 2015; Yuan \& Narayan 2014), which makes the mass accretion rate in the accretion flow decreases with decreasing radius, following $\mdot_{\rm in}(r)=\mdot_{\rm in}(R_{\rm out})\ (R/R_{\rm out})^s$, where index $s$ is the wind parameter. Following numerical simulations of non-radiative accretion flow (see Yuan, Wu \& Bu 2012 for the review of these simulations) we adopt $s\approx 0.4$ for the typical ADAF. With the increase of  accretion rate, radiative cooling becomes more and more important. Consequently the wind strength may become weaker. We thus gradually reduce parameter $s$ (down to $0.1$) for high-$\mdot$ ADAF and Type-I LHAF. For the two-phase accretion flow we fix $s=0.1$.

The second is the viscosity parameter $\alpha$. It is now widely accepted that magnetorotational instability is the mechanism to transfer angular momentum (Balbus \& Hawley 1991, 1998). The value of $\alpha$, however, is very diverse and not constrained. For example, Hawley et al. (2011) obtained $\alpha \sim 0.01-0.003$, Penna et al. (2013) obtained $\alpha\sim 0.05-0.2$, while Bai \& Stone (2013) found  that $\alpha$ can even be larger than unity. Likely the value of $\alpha$ mainly depends on the magnitude of the net flux of initial magnetic field (Pessah et al. 2007), i.e. $\alpha$ will be significantly larger when the net flux is nonzero. According to this result, the $\alpha$ value can be different in different sources, if the initial net magnetic flux contained in the accreting material from the companion star is different.

The third is the plasma $\beta$ parameter, defined as the ratio of total pressure to the magnetic pressure. It constrains the strength of magnetic fields. We simply set $\beta=10$ throughout this paper, following numerical simulations of accretion flows (Yuan \& Narayan 2014).

The forth is the turbulent dissipation parameter $\delta$, which characterizes the fraction of turbulent viscous heating that goes into electrons directly. Works have been done to estimate the values of $\delta$, by considering magnetic reconnection (Bisnovatyi-Kogan \& Lovelace 1997; Quataert \& Gruzinov 1999; Ding et al. 2010), MHD turbulence (Quataert 1998; Blackman 1999; Lehe et al. 2009), or dissipation of pressure anisotropy (Sharma et al. 2007; Sironi \& Narayan 2015). We now have the consensus that $\delta\sim 0.1-0.5$ but its exact value remains uncertain. From the observational side, by modelling the extremely dim supermassive BH, Sgr A*, Yuan et al. (2003) found that $\delta\sim 0.5$; while modelling to more luminous sources obtained a smaller value, $\delta\sim 0.1$ (Yu et al. 2011; Liu \& Wu 2013). Since the sources we will deal with in this work are much more luminous than Sgr A*, we adopt $\delta=0.1$. We will argue later in \S3.1 that $\delta=0.1$ is also favored in order to explain the radio/X-ray correlation.

The fifth is a parameter that describes the coupling between the jet and the accretion flow. It is defined as the fraction of the mass accretion rate that goes into the jet, i.e.,
\be\eta_{\rm jet} = {\mdot_{\rm jet}\over\mdot_{\rm in}(5 R_{\rm s})}.\ee
In literature, sometimes $\eta_{\rm jet}$ is assumed to be a constant, independent of $\mdot_{\rm in}(5 R_{\rm s})$, perhaps for the reason of simplicity. Most MHD numerical simulations of jet formation have neglected radiation in hot accretion flow thus are scale-free to accretion rate. Physically, when the accretion rate is very low, radiation is not important and will not affect the dynamics of jet formation. In this case, a constant $\eta_{\rm jet}$ is perhaps a good assumption. However, at higher accretion rates, radiation plays a more and more important role in the dynamics so $\eta_{\rm jet}$ may no long be independent of the accretion rate. Given the uncertainties, we in our model set $\eta_{\rm jet}$ as a free parameter. By fitting the observed correlation, we find that $\eta_{\rm jet}$ decreases with increasing accretion rate (cf. Fig.\ \ref{fig:eta}. See also YCN05). This is consistent with the following two observational facts. One is that jet is suppressed in the soft state which has higher accretion rate compared with the hard state. Another is that the degree of radio-loudness of AGNs decreases systematically with increasing Eddington ratio (Fig. 10 in Ho 2008).

\section{Modelling results}
\subsection{General description to the method}
We note that, when $L_X\la 3\times 10^{34}\ {\rm erg~s^{-1}}$,  both the `original' and the `hybrid' correlations have $p\approx 0.6$ (Coriat et al. 2011. Cf. Fig.\ \ref{fig:rx}). We first try to explain this part of correlation. The methodology is that we calculate the emission from the hot accretion flow, which dominates $L_{\rm X}$, at various accretion rate $\mdot_{\rm in}$ and that from the jet, which dominates $L_{\rm R}$, at different $\mdot_{\rm jet}$. The value of $\eta_{\rm jet}$ is adjusted to satisfy the correlation of this part. We emphasize that, the $\eta_{\rm jet}$ -- $\dot{M}(5R_{\rm s})$ relationship is different for the two type of sources, whose viscosity parameters are different. The $\eta_{\rm jet}$ -- $\dot{M}(5R_{\rm s})$ relationship can be roughly described by power-law forms, as shown by the blue (for the `original' correlation sources) and green (for the `hybrid' correlation sources) solid lines in Fig.\ \ref{fig:eta}. We then extrapolate these two power-law fitting functions to higher accretion rates and calculate the corresponding $L_{\rm R}$ and $L_{\rm X}$ to see whether we can explain the radio/X-ray correlations above $L_X\sim  3\times 10^{34}\ {\rm erg~s^{-1}}$ shown in Fig.\ \ref{fig:rx}.

\begin{figure}
\centerline{\includegraphics[width=8.5cm]{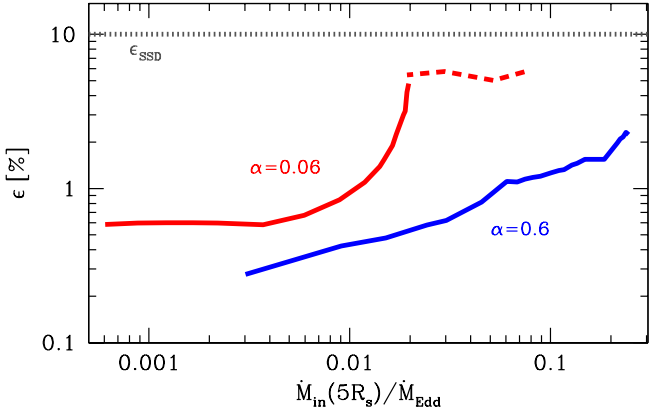}}
\caption{The radiative efficiency of hot accretion flow defined as $\epsilon=L_{\rm tot}/\mdot_{\rm in}(5R_{\rm s}) c^2$, with $L_{\rm tot}$ being the total emission from the hot accretion flow. The red and blue curves are for $\alpha=0.06$ and $0.6$, respectively. The red dashed curve is for the two-phase accretion flow.  The radiative efficiency of SSD is also shown in this plot as grey dotted curve.}
\label{fig:eff}
\end{figure}

\begin{figure}
\centerline{\includegraphics[width=8.5cm]{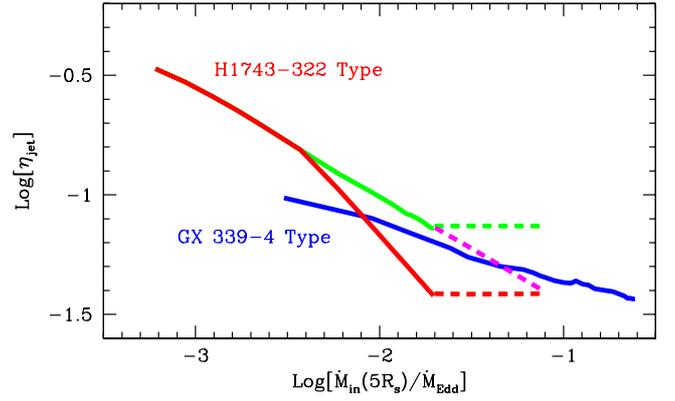}}
\caption{The $\eta_{\rm jet}$ as a function of accretion rate $\mdot_{\rm in}(5 R_{\rm s})$. The colour and type of each line are the same to those in Fig.\ \ref{fig:rx}. Note that the red and green lines are superimposed at $\mdot_{\rm in}(5R_{\rm s})\la 10^{-2.4}\ \medd$.}
\label{fig:eta}
\end{figure}

Before we introduce our results in detail, we first do some simple estimations, which is useful to understand the results presented in \S3.2 and \S3.3. We refer the readers to Fig. 1 of Xie \& Yuan (2012). This figure shows the radiative efficiency of hot accretion flow as a function of the accretion rate. Because this figure is crucial to our present work, we reproduce this figure in  Fig.\ \ref{fig:eff}, but with some modifications.  One change is the value of $\alpha$. In Xie \& Yuan (2012), $\alpha=0.1$, now we have $\alpha=0.06$ and 0.6. The second is that now the wind parameter $s$ is not a constant but is assumed to vary with the accretion rate. In addition, here we only adopt $\delta=0.1$.

Let us focus on the two curves in Xie \& Yuan (2012) corresponding to $\delta=0.5$ and $\delta=0.1$. Each curve can be divided into three branches, with their boundary at $\mdot_{\rm in} (5R_{\rm s})\sim 0.01\medd$ and $\mdot_{\rm in} (5R_{\rm s})\sim 0.004\medd$ ($\medd\equiv 10\ L_{\rm Edd}/c^2$ is the Eddington accretion rate)\footnote{Note that Xie \& Yuan (2012) used the net accretion rate $\mdot_{\rm net}$ at the event horizon. $\mdot_{\rm in} (5R_{\rm s})\approx 1.9\ \mdot_{\rm net}$, for the wind parameter $s=0.4$ adopted by Xie \& Yuan (2012).}. Both curves can be described by a flat and very steep power-law forms at the high-$\mdot_{\rm in}$ and middle-$\mdot_{\rm in}$ branches, respectively. For the low-$\mdot_{\rm in}$ branch, the $\delta=0.1$ curve can also be roughly fitted by a power-law form, while the $\delta=0.5$ curve is more curved.

To explain the `original' correlation like GX 339-4, which is described by a single power-law, it is natural to expect that the radiative efficiency of the hot accretion flow is also a single power-law, and the produced luminosity can cover the whole range of $L_{\rm X}$, up to $4\times 10^{37}~\ergs$ (refer to Fig. \ref{fig:rx} of this paper). In this sense, the $\delta=0.1$ branch is more promising than the $\delta=0.5$ one, since the low-$\mdot_{\rm in}$ branch is a single power-law while the $\delta=0.5$ branch is not. Of course, we also need that the low-$\mdot_{\rm in}$ branch can produce  $L_{\rm X}$ as high as $4\times 10^{37}~\ergs$.

To explain the `hybrid' correlation like H1743--322, it is interesting to note that the three branches of the radiative efficiency curve may correspond to the three branches of the `hybrid' correlations, namely the $p=0.6$ correlation at low $L_{\rm X}$, the transition regime at the middle $L_{\rm X}$, and the $p=1.4$ (or $p=0.98$) correlation at high $L_{\rm X}$, respectively. For example, the steep power-law efficiency curve means that with small increase of $\mdot_{\rm in}$, $L_{\rm X}$ will increase rapidly. This corresponds to the transition regime of the hybrid correlation. Again, the $\delta=0.1$ curve in Xie \& Yuan (2012) is better than the $\delta=0.5$ one, since the steep power-law branch covers a wider range of radiative efficiency which is more suitable to explain the broad (about one order of magnitude in $L_{\rm X}$) transition regime of the hybrid correlation. Below we present our detailed results.

\subsection{The `original' correlation}

We adopt $\alpha=0.6$ and $\delta=0.1$. The solid blue line in Fig.\ \ref{fig:rx} shows the numerical result of the correlation (Note that below $2.5\times 10^{34}\ \ergs$, the curves with different colors are superimposed with each other.). The observed `original' $p=0.6$ correlation is reproduced, except at $L_X\ga 10^{37}\ergs$ (but see our discussion at \S4). The corresponding solid blue line in Figs.\ \ref{fig:eff} and \ref{fig:eta} show, respectively, the radiative efficiency $\epsilon$ and parameter $\eta_{\rm jet}$ as a function of $\mdot_{\rm in}(5R_{\rm s})$. The typical value of $\eta_{\rm jet}(\sim 10\%$) is consistent with what we have obtained in MHD numerical simulations of `disk-jet' (e.g., Hawley \& Krolik 2006; Ohsuga \& Mineshige 2011; Yuan et al. 2015).

The values of $\alpha$ is large compared with the value usually adopted in literature, but it is well within the reasonable range obtained from MHD numerical simulations (refer to \S2). The reason for this choice is that, the critical accretion rate of ADAF, $\mdot_{\rm cr, ADAF}\approx (0.1-0.3)\alpha^2\mdot_{\rm Edd}\propto \alpha^2$ (Yuan\& Narayan 2014), will be large. Here the critical accretion rate measures the net accretion rate $\mdot_{\rm in}$\footnote{The largest $\mdot_{\rm in}(5R_{\rm s})$ in the solid blue line in Fig.\ \ref{fig:eta} is slightly larger than $0.3\alpha^2\mdot_{\rm Edd}$. This is because wind parameter $s$ adopted here is smaller than that in Yuan \& Narayan (2014).}. Therefore the luminosity from the ADAF can be high enough to roughly cover the full range of $L_{\rm X}$ of the original correlation. This, together with the fact that the radiative efficiency of an ADAF can be described by a single power-law in the case of $\delta=0.1$ (cf. Fig.\ \ref{fig:eff}), explains why we can produce (nearly) a single power-law radio/X-ray correlation shown by the solid blue curve in Fig.\ \ref{fig:rx}.

\subsection{The `hybrid' correlation}

We here choose $\alpha=0.06$ and $\delta=0.1$ for this type of sources. We have tried several forms of $\eta_{\rm jet}$ -- $\mdot_{\rm in}(5R_{\rm s})$ functions, as shown by the (solid and dashed) red, green, and magenta lines in Fig.\ \ref{fig:eta}. The radio/X-ray correlation results are shown by the corresponding lines in Fig.\ \ref{fig:rx}. Below we will discuss these fitting results one by one.

First, let's look at the solid green lines in Figs.\ \ref{fig:rx}\& \ref{fig:eta}. The lines below and above $\sim 2\times10^{35}\ \ergs$ in Fig.\ \ref{fig:rx} correspond to the ADAF and type-I LHAF, respectively.  In Fig.\ \ref{fig:eta} the solid green line is similar to the solid blue line (denoting the `original' correlation) in the sense that both are straight lines (i.e., power-law function), but the radio/X-ray correlation results shown in Fig.\ \ref{fig:rx} (solid blue and green lines) are different. The blue line is a single power-law while the green line becomes almost flat at $L_X\ga 2\times 10^{35}\ergs$. This is because the dependence of radiative efficiency on the accretion rate is different for ADAF and type-I LHAF, as we have explained in \S3.1 (cf. Fig.\ \ref{fig:eff}).

We have extrapolated the solid green line to higher $\mdot_{\rm in}(5R_{\rm s})$, i.e. the two-phase accretion mode, as shown by the dashed magenta line in Fig.\ \ref{fig:eta}. We then obtain a radio/X-ray correlation with correlation index $p\approx 1$, as shown by the dashed magenta line in Fig. \ref{fig:rx}. This is in agreement with the correlation found by Gallo et al. (2012).

However, theoretical considerations suggest that the $\eta_{\rm jet}$ function may not extend as a simple power-law form as shown by the dashed magenta line in Fig.\ \ref{fig:eta}. This is because the accretion flow in the regime of dashed magenta line is in two-phase regime. For such kind of accretion flow, it is expected that, as the total (cold and hot) accretion rate increases, more fraction of the whole accreting gas will be in the cold phase. However, in this paper the accretion rate at the two-phase regime only takes the hot gas into account. Numerical MHD simulations of jet formation suggest that one important factor to determine the strength of jet is the magnitude of magnetic flux accumulated in the BH horizon or the innermost region of the accretion flow (\S3.3 in Yuan \& Narayan 2014), which is carried in by both the hot gas and the cold clumps. So for a two-phase accretion flow, with the increase of $\mdot_{\rm in}(5R_{\rm s})$, $\eta_{\rm jet}$ is more likely to be larger and larger, compared with the corresponding value shown by the dashed magenta line in Fig. \ref{fig:eta}. For simplicity we assume a constant $\eta_{\rm jet}$, as shown by the dashed green line in Fig. \ref{fig:eta}. This results in a radio/X-ray correlation with $p\approx 1.4$, as shown by the green dashed curve in Fig. \ref{fig:rx}. This is consistent with that obtained by Coriat et al. (2011).

The solid and dashed green curves can roughly reproduce three branches of the `hybrid' correlation with correct power-law index $p$, namely the low branch of $p\approx 0.6$, the transition regime with  $p\sim 0$, and the steep branch of $p\approx1.4$. However, the fitting to the flat correlation branch, with $L_X\ga \sim 3\times 10^{34}\ \ergs$, is clearly not satisfactory since the model overpredicts the $L_{\rm R}$ by a factor of $\la 2.5$.

To improve the fitting, we test another $\eta_{\rm jet}--\mdot_{\rm in}(5R_{\rm s})$ function, as shown by the (solid and dashed) red lines in Fig. \ref{fig:eta}. Compared with the green lines, the solid red line decreases faster with increasing $\dot{M}_{\rm in}(5R_{\rm s})$. We also assume $\eta_{\rm jet}$ to be a constant at the two-phase accretion flow regime, i.e., $\mdot_{\rm in}(5R_{\rm s})\ga 0.02\ \medd$. The corresponding fitting results are shown by the (solid and dashed) red lines in Fig. \ref{fig:rx}. We see from Fig. \ref{fig:rx} that now the model can explain both the $p\sim 0$ and the $p=1.4$ correlations quantitatively well.

The physical reason for the dashed red line in Fig. \ref{fig:eta} is the same with the dashed green line. Now the question is whether the solid red line physical? Or, given the uncertainty in jet formation theory, what information can we learn from this result? We note that the deviation of the solid red line from the solid green line in Fig.\ \ref{fig:eta} begins from a relatively large $\mdot_{\rm in}$ close to $\mdot_{\rm cr,ADAF}$. This suggests that, as the accretion flow enters into the LHAF regime from the ADAF one, the radiative cooling becomes more important in affecting the dynamics of the accretion flow, including the jet formation process. This hypothesis can be tested in future by radiative MHD simulations of jet formation.

\section{Summary and Discussion}

In term of the radio/X-ray correlation, $L_R\propto L_X^p$, two types of sources have been observed. In the `original' type, the correlation follows a simple correlation which can be described by a single power-law function, with the power-law index $p\approx 0.6$. For the second type of sources, their correlation is `hybrid', which is described by three power-law functions with different index $p$ at the different luminosity regimes. In this paper, we try to explain these two types of correlation based on the coupled accretion--jet model of YCN05. This model has been successfully applied to model the multi-waveband spectrum of the hard state of BHBs and low-luminosity AGNs, and has explained the `original' correlation (Yuan \& Cui 2005; Yuan \& Narayan 2014).

In this model, the jet and the hot accretion flow are responsible for $L_{\rm R}$ and $L_{\rm X}$, respectively. While $L_{\rm R}$ always follows a power-law function of the mass loss rate in the jet, the dependence of $L_{\rm X}$ on the accretion rate is much more complicated. Depending on the specific mode of the hot accretion, namely ADAF, type-I LHAF, and two-phase accretion flow, the relationship between the radiative efficiency and the accretion rate varies. This results in different values of $p$ in the $L_{\rm R} -- L_{\rm X}$ correlation. The key assumption in this work is that, sources that always follow the `original' correlation have a large viscosity parameter $\alpha$; therefore the critical rate of ADAF is large and the accretion flow can always stay in the ADAF regime throughout the whole observed range of $L_{\rm X}$. This explains why their correlation can be described by a single power-law form. On the other hand, the sources following the `hybrid' correlation have a small $\alpha$ so the critical rate of ADAF is small. In this case, with the increase of accretion rate, their accretion modes change from ADAF to type-I LHAF, then to two-phase accretion flow. This explains why there are three branches of correlation for these sources. Our model indicates that the flat $p\sim 0$ branch of the `hybrid' correlation is actually X-ray bright (due to a quick enhancement in the radiative efficiency as $\mdot_{\rm in}$ increases) instead of radio faint. As suggested by MHD numerical simulations, the physical reason for the different value of $\alpha$ may be because of the difference in the net magnetic flux carried by the accreted material in the two types of sources. Unfortunately, the net magnetic flux of the accreting gas is unclear either observationally or theoretically. One possibility is that it may relate to the magnetic field configuration of the companion star.

During the modelling, we find that the ratio between the mass-loss rate and the accretion rate $\eta_{\rm jet}$ is not a constant of the accretion rate $\mdot_{\rm in}$. Rather, in order to explain the observed radio/X-ray correlations, $\eta_{\rm jet}$ must decrease with increasing $\mdot_{\rm in}$. Observationally, this is consistent with the facts that jets are present only in the hard state, and that the radio loudness of AGNs decreases with increasing Eddington rate. Theoretically, this indicates that radiation plays a certain role in affecting the jet formation. Detailed radiative MHD simulations of jet formation are required to test this hypothesis.

Although a large $\alpha=0.6$ is adopted when we explain the `original' correlation, we still fail to reproduce the highest $L_{\rm X}$ (refer fig.\ \ref{fig:rx}). We note that GX 339--4 has almost the highest hard state $L_{\rm X}$ among all BH binaries detected. One comment is that the BH spin of this source may be large, $a>0.9$ (e.g. Miller et al. 2004; Reis et al. 2008; Yamada et al. 2009), while our calculations are for accretion on to a Schwarzschild BH. If a large BH spin were taken into account, we would produce a higher $L_{\rm X}$ since the radiative efficiency of accretion flow around spinning BH will be higher.

One special source is Cyg X--1, where the BH accretes material from the wind of the high-mass companion. Its radio emission suffers additional free--free absorption by the stellar wind. After corrections of this absorption, the intrinsic radio/X-ray correlation of this source is likely $p\approx 1.4$ (fig. 11 in Zdziarski et al. 2011). One caveat here is that some uncertainties exists in the correction of free-free absorption. If the $p\approx 1.4$ result is correct, under the scenario proposed in this work, it is likely that the viscosity parameter $\alpha$ of Cyg X-1 is moderately small, and the hard state of this source is described by the two-phase accretion.

\section*{Acknowledgments}

We thank M. Coriat and S. Corbel for providing us the observational data, and appreciate A.~A. Zdziarski and the referee for comments. This work was supported in part by the Natural Science Foundation of China (grants 11133005, 11203057 and 11573051), the National Basic Research Program of China (973 Program, grant 2014CB845800), the Strategic Priority Research Program `The Emergence of Cosmological Structures' of CAS (grant XDB09000000), and the CAS/SAFEA International Partnership Program for Creative Research Teams.

\bsp	
\label{lastpage}
\end{document}